# Ergodic Capacity of Composite Fading Channels in Cognitive Radios with Series Formula for Product of *κ-μ* and *α-μ* Fading Distributions

He Huang, *et.al*

Beijing University of Posts and Telecommunications, Beijing 100876, China

*Abstract*—In this study, product model of two independent and non-identically distributed (i.n.i.d.) random variables (RVs) for *κ-μ* fading distribution and *α-μ* fading distribution is considered. The method of the product for RVs has been broadly applied in a large number of communications fields, such as cascaded fading channels, multiple input multiple output (MIMO) systems, radar communications and cognitive radio networks (CRs). The novel exact series representations for the product of two i.n.i.d. fading distributions *κ-μ* and *α-μ* are derived instead of Fox H-function to solve the problem that Fox H-function with multiple RVs cannot be implemented in professional mathematical software packages as MATHEMATICA and MAPLE. Exact close-form expressions of probability density function (PDF) and cumulative distribution function (CDF) for proposed models are deduced to represent the provided product expressions and generalized composite multipath shadowing models. Furthermore, ergodic channel capacity (ECC) representations are obtained to measure maximum fading channel capacity in CRs. At last, these analytical results are validated with Monte Carlo simulations to evaluate spectrum efficiency over generalized composite shadowing fading scenarios in CRs.

## I. Introduction

### A. Background

In cognitive radios networks(CRs) wireless communication systems, there have been a number of important studies with regard to the product of random variables (RVs), which is applied to optimize spectrum efficiency and evaluate performance of communication systems [1]-[4]. Different models of the product with multiple RVs have been universally investigated to model high-resolution synthetic radar clutter [5], obtain the channel gain and keyhole effects over cascaded fading channels in multiple-input multiple-output (MIMO) [6], [7], and combine long term fading and short term fading in vehicle-to-vehicle communications and body area networks [8]-[12]. Meanwhile, performance analysis of channel capacity has been regarded as the essential tool to improve spectrum efficiency in CRs recently. Hence, channel capacity statistics analysis with the product of RVs has generally raised broad interests in wireless communication.

The composite fading distributions describe the common fading phenomena for long term fading and short term fading in non-stationary stochastic environments [8], [13], it is more practical to jointly represent the large-scale and small-scale fading characteristics in realistic wireless fading channels. The short term fading can be observed within short distance with few wavelengths, such as *κ-μ*, *η-μ*, *α-μ*, Rayleigh, Nakagami-*m* and Weibull [14], [15], and in general the long term fading is described by lognormal distribution which can be replaced by *α-μ* distribution (consists of special cases as gamma, negative exponential, Nakagami-*m*, Weibull, one-sided Gaussian and Rayleigh) [8], [9]. The *κ-μ* distribution is a generalized small-scale line-of-sight (LOS) multipath fading distribution. With the aid of circularly symmetric random variables, the *κ-μ* distribution can model the scattering cluster in homogeneous communication environments and it includes Rice ($\kappa=k$, $\mu=1$), Rayleigh ($\kappa=0$, $\mu=1$), Nakagami-*m* ($\kappa=0$, $\mu=m$) and one-sided Gaussian distributions ($\kappa=0$, $\mu=0.5$) for two fading parameters $\kappa$ and $\mu$ [3], [4], [15]. Moreover, the *α-μ* distribution is general fading model which describes the signal is made up of multipath waves clusters in non-linear fading condition, and it can better accommodate statistical characteristic variations for non-linear signal because of its flexibility. Besides, a large number of field testing results show that gamma distribution can accurately approximate the lognormal distribution to solve the problem that the close-form expressions of lognormal process is hard to obtain [8], [9], [16].

In addition, in wireless communication the composite fading distributions have been investigated to show multipath and shadow phenomena for fading channels, exact and unified versatile expressions with cognitive radios techniques and performance analysis of channel capacity have been analyzed over the generalized composite shadowing models, such as *κ-μ*/gamma, *η-μ*/gamma and *α-μ*/gamma. The probability density function (PDF) and the cumulative distribution function (CDF) are derived to help estimate outage probability (OP) and average bit error probability (ABEP) with diversity reception [17]-[20]. At the same time, channel capacity has been defined as a good research standard for spectrum efficiency when at the receiver the channel state information (CSI) is available. The ergodic channel capacity (ECC) is an important measurement because it effectively measures the maximum capacity with optimal rate adaptation (ORA) under Gaussian scenario, and novel exact close-form expressions of ECC have been derived over different generalized fading channels to evaluate the spectral efficiency in fading scenarios [21]-[24].

### B. Main contributions

Although there are many papers about the product method for fading models, such as, the PDF and CDF for the product model of multiple *α-μ* variates have been investigated in [8], [9] by calculating finite sum of hypergeometric functions. The MATHEMATICA routine for the product of Nakagami-*m* distributions has been proposed to deal the problem that Fox H-function could not overall be implemented in MATHEMATICA and MAPLE [25]. The CDF and PDF for product of three *α-μ* variates are derived by using Meijer G-function [26]. Furthermore, in [12] the general structure for PDF and CDF with Fox H-function have been derived as [12, Eq. (1), Eq. (15) and Eq. (30)] with residues calculus, and the outage capacity [12, Eq. (27)] and detection probability in a UHF (Ultra High Frequency) RFID (Radio Frequency Identification) system [12, Eq. (28)] have been derived by the form of integrals, however, the exact close-form series expressions for *κ-μ*/*α-μ* and *η-μ*/*α-μ* with generalized hypergeometric function have not been derived in [12, Table III and Table IV] and [12, Table III and Table IV] are just special series cases of product of fading models with Gauss hypergeometric function [24, Eq. (15.1.1)]. At the same time, in [12] the conclusion states that 'there are maybe other more exact series expressions but with more algebraic manipulations' for the unknown product models, besides, the single-variable and multi-variable Fox H-function are not yet available in MATHEMATICA and MAPLE.

Motivated by above, as series provide more precise evaluations and compute more quickly in traditional computer, and there are no studies related to exact close-form series expressions for the product of two i.n.i.d. RVs with $\kappa$-$\mu$ and $\alpha$-$\mu$ distribution in the open literature. Hence, in this paper, the generalized product models $\kappa$-$\mu$/$\alpha$-$\mu$ have been proposed to approximate the generalized composite multipath shadowing channels, and it is the first time that ECC has been derived over the proposed generalized structure. The main contributions of this paper are summarized as follows:

· This paper provides exact series product model $\kappa$-$\mu$/$\alpha$-$\mu$ for two i.n.i.d. RVs with fading distribution by employing generalized hypergeometric function instead of Fox H-function [12], which describes the intertwined effects of line-of-sight (LOS) short-term fading $\kappa$-$\mu$ distribution and long-term fading distribution that is substituted by $\alpha$-$\mu$ distribution. The provided novel model can consider special cases, for example, Rice/$\alpha$-$\mu$, Rayleigh/$\alpha$-$\mu$, Nakagami-$m$/$\alpha$-$\mu$ and one-sided Gaussian/$\alpha$-$\mu$ (where $\alpha$-$\mu$ distribution consists of gamma, Nakagami-$m$, negative exponential, Weibull, one-sided Gaussian and Rayleigh distribution).

· Novel exact analytical series expressions for the product model of two i.n.i.d. RVs $X$ for $\kappa$-$\mu$ distribution and $Y$ for $\alpha$-$\mu$ distribution have been obtained with the help of the fading envelope, the representations for PDF and CDF of $\kappa$-$\mu$/$\alpha$-$\mu$ are derived to represent the combined product model with short term fading and long term fading.

· The ECC expressions for composite multipath-shadowing fading channels $\kappa$-$\mu$/$\alpha$-$\mu$ are derived to measure the spectrum efficiency in CRs.

This paper is organized as follows. In Section II, $\kappa$-$\mu$ fading distribution and $\alpha$-$\mu$ fading distribution have been are listed. In Section III, the product model of $\kappa$-$\mu$ and $\alpha$-$\mu$ fading distributions are derived. In Section IV, ECC of the composite fading models are derived. In Section V numerical results and application are presented. Finally, Section VI concludes this paper.

## II. FADING DISTRIBUTIONS

The $\kappa$-$\mu$ fading distribution model small-scale LOS signals propagated in homogeneous scattering environments [15]. For fading signal with envelope $R$, normalized envelope $P(P = R/\hat{r})$, the root-mean-square (rms) of $R$ is $\hat{r}$ ($\hat{r} = \sqrt{E(R^2)}$, $E(\cdot)$ denotes the expectation), the envelope PDF $f_P(\rho)$ can be written as

$$f_P(\rho) = \frac{2\mu(1+\kappa)^{\frac{\mu+1}{2}}}{\kappa^{\frac{\mu-1}{2}} e^{\kappa\mu}} \rho^\mu e^{-\mu(1+\kappa)\rho^2} I_{\mu-1}(2\mu\sqrt{\kappa(1+\kappa)}\rho) \quad (1)$$

where $\kappa(\kappa>0)$ is the ratio between the total power of the dominant components and the total power of the scattered waves, $\mu(\mu>0)$ is the number of multipath clusters, $I_\nu(\cdot)$ is the modified Bessel function of the first kind with the order $\nu$. For fading signal with power $W=R^2$ and normalized power $\Omega=W/\bar{w}$ ($\bar{w}=E(W)$), the power PDF $f_\Omega(w)$ is given by

$$f_\Omega(\omega) = \frac{\mu(1+\kappa)^{\frac{\mu+1}{2}}}{\kappa^{\frac{\mu-1}{2}} e^{\kappa\mu}} \omega^{\frac{\mu-1}{2}} e^{-\mu(1+\kappa)\omega^2} I_{\mu-1}[2\mu\sqrt{\kappa(1+\kappa)}\omega] \quad (2)$$

The rms of $\kappa$-$\mu$ distribution is given by

$$\hat{r}^{\kappa-\mu} = \sqrt{E(R^2)} = \frac{\bar{r}\Gamma(\mu)(1+\kappa)^{1/2}\mu^{1/2}e^{\kappa\mu}}{\Gamma(\mu+\frac{1}{2})_1F_1(\mu+1/2,\mu,\kappa\mu)} \quad (3)$$

where $\bar{r} = E(R)$, $\Gamma(\cdot)$ is the gamma function.

The $\alpha$-$\mu$ fading distribution is a non-linear physical fading model with fading parameters $\alpha$ and $\mu$ [8], [9], for the fading signal envelope $R$, the envelope PDF can be expressed as

$$f_R(r) = \frac{\alpha\mu^\mu}{\Gamma(\mu)} \frac{r^{\alpha\mu-1}}{\hat{r}^{\alpha\mu}} e^{-\mu\frac{r^\alpha}{\hat{r}^\alpha}} \quad (4)$$

where the parameter $\alpha(\alpha>0)$ represents the nonlinearity of the propagation medium and the parameter $\mu(\mu>0)$ is the number of multipath clusters, the $\alpha$-rms of $\alpha$-$\mu$ distribution is given by

$$\hat{r}^{\alpha-\mu} = \sqrt[\alpha]{E(R^\alpha)} = \frac{\bar{r}\mu^{\frac{1}{\alpha}}\Gamma(\mu)}{\Gamma(\mu+\frac{1}{\alpha})} \quad (5)$$

## III. PRODUCT MODEL OF I.N.I.D. FADING DISTRIBUTIONS

### A. Product model of two i.n.i.d. RVs

Assuming two i.n.i.d. positive RVs $X$ for $\kappa$-$\mu$ fading distribution and $Y$ for $\alpha$-$\mu$ fading distribution respectively, let $Z = X \times Y$, with the help of the principles of probability theory and statistics, the PDF for RV $Z$ is given by [8]

$$f_Z(z) = \int_0^\infty f_Y(y) f_{Z|Y}(z|y) dy \quad (6)$$

where $f_Z(\cdot)$ and $f_Y(\cdot)$ are the PDFs of $Z$ and $Y$ respectively, $f_{Z|Y}(\cdot)$ is the conditional PDF of $Z/Y$ and it can be given by

$$F_{Z|Y}(\frac{z}{y}) = \frac{1}{y} f_X(\frac{z}{y}) \quad (7)$$

where $F_{Z|Y}(z/y)$ denotes the CDF of $Z/Y$, $X=Z/Y$, the PDF of $Z$ with $\kappa$-$\mu$ distribution for RV $X$ and $\alpha$-$\mu$ distribution for RV $Y$ can be expressed as

$$f_Z(z;\bar{x},\bar{y}) = \int_0^\infty \frac{1}{y} f_{Z|Y}^{\kappa-\mu}(z|y;\bar{x}) f_Y^{\alpha-\mu}(y;\bar{y}) dy \quad (8)$$

where $\bar{x} = E(X)$ and $\bar{y} = E(X)$. For $\kappa$-$\mu$ distribution of RV $X$ in (8), with the aid of (1)-(3), the conditional PDF in (8) can be expressed as

$$f_{R|x}^{\kappa-\mu}(r_x) = A_1 \frac{r_x^{\mu_x}}{\hat{r}_x^{\mu_x+1}} e^{-\mu_x(1+\kappa_x)\frac{r_x^2}{\hat{r}_x^2}} I_{\mu_x-1}(2\mu_x\sqrt{\kappa_x(1+\kappa_x)}\frac{r_x}{\hat{r}_x}) \quad (9)$$

where

$$A_1 = \frac{2\mu_x(1+\kappa_x)^{\frac{\mu_x+1}{2}}}{\kappa_x^{\frac{\mu_x-1}{2}} e^{\kappa_x\mu_x}}, \quad \hat{r}_x = \frac{\bar{r}_x\Gamma(\mu_x)(1+\kappa_x)^{1/2}\mu_x^{1/2}e^{\kappa_x\mu_x}}{\Gamma(\mu_x+\frac{1}{2})_1F_1(\mu_x+1/2,\mu_x,\kappa_x\mu_x)} \quad (10)$$

where $\bar{r}_x = E(R_x)$. With the help of (4) and (5), the envelope PDF for $\alpha$-$\mu$ distribution of RV $Y$ in (8) is given by

$$f_{R|y}^{\alpha-\mu}(r) = \frac{\alpha_y\mu_y^{\mu_y}}{\Gamma(\mu_y)} \frac{r_y^{\alpha_y\mu_y-1}}{\hat{r}_y^{\alpha_y\mu_y}} e^{-\mu_y\frac{r_y^{\alpha_y}}{\hat{r}_y^{\alpha_y}}} \quad (11)$$

where

$$\hat{r}_y = \sqrt[\alpha_y]{E(R^{\alpha_y})} = \frac{\bar{r}_y \mu_y^{\frac{1}{\alpha_y}} \Gamma(\mu_y)}{\Gamma(\mu_y + \frac{1}{\alpha_y})} \tag{12}$$

in (12) $\bar{r}_y = E(R_y)$. By substituting (9)-(12) into (8), the product PDF for $Z$ can be obtained as

$$f_Z(z) = A_1 A_2 \underbrace{\int_0^\infty y^{-\mu_x + \alpha_y \mu_y - 2} e^{-\frac{\mu_x(1+\kappa_x)z^2}{\hat{r}_x^2 y^2} - \frac{\mu_y}{\hat{r}_y^{\alpha_y}} y^{\alpha_y}} I_{\mu_x-1}\left(\frac{2\mu_x\sqrt{\kappa_x(1+\kappa_x)}}{\hat{r}_x y} z\right) dy}_{C_1} \tag{13}$$

where

$$A_2 = \frac{\alpha_y \mu_y^{\mu_y} z^{\mu_x}}{\Gamma(\mu_y) \hat{r}_y^{\alpha_y \mu_y} \hat{r}_x^{\mu_x+1}} \tag{14}$$

### B. Product PDF

Having established the PDF of $Z$ for $\kappa$-$\mu$ distribution of RV $X$ with parameters $\kappa_x, \mu_x, \hat{r}_x$, and $\alpha$-$\mu$ distribution of RV $Y$ with $\alpha_y, \mu_y, \hat{r}_y$, the PDF can be derived as Theorem 1.

*Theorem* 1: the PDF of the product of i.n.i.d. RV $X$ and $Y$ ($Z = X \times Y$) can be expressed as (15), where RV $X$ is $\kappa$-$\mu$ distribution and RV $Y$ is $\alpha$-$\mu$ distribution,
in (15) where

$$\varphi(k, \kappa_x, \mu_x, \alpha_y, \mu_y) = \frac{A_1 A_2 A_3}{2\Gamma(k+1)\Gamma(k+\mu_x)} \left(\frac{\mu_x^2 \kappa_x (1+\kappa_x)}{\hat{r}_x^2}\right)^k$$

$$z^{2k} \frac{2}{\alpha_y B_1^{k+\mu_x} B_2^{\frac{\alpha_y \mu_y}{2}}},$$

$$A_3 = \left(\frac{\mu_x \kappa_x^{\frac{1}{2}} (1+\kappa_x)^{\frac{1}{2}} z}{\hat{r}_x}\right)^{\mu_x-1}, B_1 = \frac{\mu_x(1+\kappa_x) z^2}{\hat{r}_x^2}, B_2 = \left(\frac{\mu_y}{\hat{r}_y^{\alpha_y}}\right)^{\frac{2}{\alpha_y}}, \tag{16}$$

$$\Xi(a,b) = \frac{b}{a}, \frac{b+1}{a}, \ldots, \frac{b+a-1}{a}, \frac{p}{q} = \frac{1}{\frac{\alpha_y}{2}}$$

in which $_mF_n()$ is the generalized hypergeometric function [27, Eq. (9.14.1)].

*Proof*: For integral $C_1$ in (13), suppose $t = y^2$, the $C_1$ is expressed as

$$C_1 = \frac{1}{2} \int_0^\infty t^{\frac{-\mu_x + \alpha_y \mu_y - 2}{2}} e^{-\frac{\mu_x(1+\kappa_x)z^2}{\hat{r}_x^2 t} - \frac{\mu_y}{\hat{r}_y^{\alpha_y}} t^{\frac{\alpha_y}{2}}} \underbrace{I_{\mu_x-1}\left(\frac{2\mu_x\sqrt{\kappa_x(1+\kappa_x)}z}{\hat{r}_x t^{\frac{1}{2}}}\right)}_{C_2} dt \tag{17}$$

where $I_\nu(z)$ can be obtained as [27, Eq. (8.445)].

$$I_\nu(z) = \sum_{k=0}^\infty \frac{1}{\Gamma(k+1)\Gamma(\nu+k+1)} \left(\frac{z}{2}\right)^{2k+\nu} \tag{18}$$

Substituting (18) into (17), the $C_2$ in (17) can be simplified as

$$C_2 = \sum_{k=0}^\infty \frac{1}{\Gamma(k+1)\Gamma(k+\mu_x)} \left(\frac{\mu_x\sqrt{\kappa_x(1+\kappa_x)}z}{\hat{r}_x t^{\frac{1}{2}}}\right)^{2k+\mu_x-1} \tag{19}$$

Then substituting (17) and (19) into (13), the PDF of $Z$ can be given by

$$f_Z(z) = \frac{A_1 A_2 A_3}{2} \sum_{k=0}^\infty \frac{1}{\Gamma(k+1)\Gamma(k+\mu_x)} \left(\frac{\mu_x^2 \kappa_x(1+\kappa_x)}{\hat{r}_x^2}\right)^k z^{2k}$$

$$\underbrace{\int_0^\infty t^{-(k+\mu_x) + \frac{\alpha_y \mu_y}{2} - 1} e^{-\frac{\mu_x(1+\kappa_x)z^2}{\hat{r}_x^2 t} - \frac{\mu_y}{\hat{r}_y^{\alpha_y}} t^{\frac{\alpha_y}{2}}} dt}_{I_1} \tag{20}$$

To the best of auhtors' knowledge, the integral $I_1$ in (20) could not be solved with any formula for the uncertainty coefficients of $t$ in exponential function, because the integral diverges as an infinite series expansion when it is applied to the exponential function. However, with the aid of [8, Eq. (12)-(15)] and [28, Eq. (2.3.2.14)], after algebraic manipulations we construct the integral (21) and let $\alpha_y/2:1 = p:q$ ($p$ and $q$ are the positive coprime integers). It is noteworthy that when $p$ or $q$ are positive even numbers, similarly, like [8], [9] and [12], we get $p$-1 for $p$ or $q$-1 for $q$. Then, the formula (20) can be derived as (15).

### C. Product CDF

*Theorem* 2: the CDF of the product of $\kappa$-$\mu$ distribution for RV $X$ and $\alpha$-$\mu$ distribution for RV $Y$ can be expressed as (22), when $X$ and $Y$ are i.n.i.d RVs. In (22) where

$$M_1 = \frac{\alpha_y \mu_y^{\mu_y}}{\Gamma(\mu_y) \hat{r}_y^{\alpha_y \mu_y} \hat{r}_x^{\mu_x+1}}, M_2 = \left(\frac{\mu_x \kappa_x^{\frac{1}{2}}(1+\kappa_x)^{\frac{1}{2}}}{\hat{r}_x}\right)^{\mu_x-1}, M_3 = \frac{\mu_x(1+\kappa_x)}{\hat{r}_x^2} \tag{23}$$

*Proof:* the CDF of $Z$ can be obtain as

$$F_Z(z) = \int f_Z(z) dz \tag{24}$$

With the help of (15), the CDF of $Z$ can be derived by substituting (15) into (24), after some algebraic simplification, the CDF of $Z$ could be derived as (22) with the aid of (16), (21) and [29, Eq. (1.16.1.1)].

### D. Composite fading channels

The composite fading characteristics can be represented as short-term distribution/long-term distribution and it will be considered as the special expressions when $\bar{x}=1$ for the product model $\kappa$-$\mu/\alpha$-$\mu$. Then the special composite fading expressions for Rice/$\alpha$-$\mu$, Rayleigh/$\alpha$-$\mu$, Nakagami-$m/\alpha$-$\mu$ and one-sided Gaussian/$\alpha$-$\mu$ will be obtained from (15) and (22).

## IV. ERGODIC CHANNEL CAPACITY OF PRODUCT MODEL

The analytical expressions of average channel capacity are derived over the proposed composite fading model $\kappa$-$\mu/\alpha$-$\mu$ under ORA to measure the maximum channel capacity. As the channel fade level has not been known, the ECC can be achieved by using variable-rate transmission relative to the channel conditions while the transmit power remains constant [30], [31]. The ECC is obtained by averaging the capacity of an additive white Gaussian noise (AWGN) channel and it can be defined by

$$\bar{C} = \frac{B}{\ln 2} \int_0^\infty \ln(1+\gamma) f_\gamma(\gamma) d\gamma \tag{25}$$

where $\bar{C}$ is the average channel capacity, $B$ is the bandwidth of channel, $\gamma$ is the instantaneous signal-to-noise ratio (SNR), $f_\gamma(\gamma)$ is the power PDF of the fading distribution.

*Theorem* 3: Under the ORA transmission policy, the ECC expression of the product model $\kappa$-$\mu/\alpha$-$\mu$ of two i.n.i.d. RVs $X$ and $Y$ is derived as (26), shown in page 4. Where $\bar{\gamma}$ denotes the average SNR.

*Proof*: with the help of (15) and [15, Eq. (9)-(11)], For the fading signal with the power $w=R^2$ and the normalized power is $w/E(w)$ from the envelope PDF in (15), the power PDF can be obtained as (27). In (25),

$$f_Z(z) = \sum_{k=0}^{\infty} \varphi(k,\kappa_x,\mu_x,\alpha_y,\mu_y)((B_1B_2)^{k+\mu_x} \sum_{i=0}^{q-1} \frac{(-1)^i}{i!} \Gamma(-\frac{p}{q}(i+k+\mu_x)+\mu_y)(B_1B_2)^i$$
$$_1F_{p+q}(1;\Xi(p,\frac{p}{q}(i+k+\mu_x)-\mu_y+1),\Xi(q,i+1);\frac{(B_1B_2)^q}{(-p)^p(-q)^q}) + \frac{\alpha_y}{2}(B_1B_2)^{\frac{\alpha_y\mu_y}{2}} \sum_{i=0}^{p-1} \frac{(-1)^i}{i!} \Gamma(k+\mu_x-\frac{q}{p}(i+\mu_y))(B_1B_2)^{\frac{\alpha_y i}{2}}$$
$$_1F_{p+q}(1;\Xi(q,\frac{q}{p}(i+\mu_y)-(k+\mu_x)+1),\Xi(p,i+1);\frac{(B_1B_2)^{\frac{\alpha_y p}{2}}}{(-p)^p(-q)^q})) \quad (15)$$

---

$$(B_1)^{\mu_x}(B_2)^{\frac{\alpha_y\mu_y}{2}} \int_0^{\infty} t^{\frac{\alpha_y\mu_y}{2}-\mu_x-1} \exp(-B_1 t^{-1} - (B_2)^{\frac{\alpha_y}{2}} t^{\frac{\alpha_y}{2}}) dt = \frac{2}{\alpha_y}((B_1B_2)^{\mu_x} \sum_{i=0}^{q-1} \frac{(-1)^i}{i!} \Gamma(-\frac{p}{q}(i+\mu_x)+\mu_y)(B_1B_2)^i$$
$$_1F_{p+q}(1;\Xi(p,\frac{p}{q}(i+\mu_x)-\mu_y+1),\Xi(q,i+1);\frac{(B_1B_2)^q}{(-p)^p(-q)^q}) + \frac{\alpha_y}{2}(B_1B_2)^{\frac{\alpha_y\mu_y}{2}} \sum_{i=0}^{p-1} \frac{(-1)^i}{i!} \Gamma(\mu_x-\frac{q}{p}(i+\mu_y))(B_1B_2)^{\frac{\alpha_y i}{2}}$$
$$_1F_{p+q}(1;\Xi(q,\frac{q}{p}(i+\mu_y)-\mu_x+1),\Xi(p,i+1);\frac{(B_1B_2)^{\frac{\alpha_y p}{2}}}{(-p)^p(-q)^q})) \quad (21)$$

---

$$F_Z(z) = \sum_{k=0}^{\infty} \frac{A_1 M_1 M_2}{2\Gamma(k+1)\Gamma(k+\mu_x)} (\frac{\mu_x^2 \kappa_x(1+\kappa_x)}{\hat{r}_x^2})^k \frac{2}{\alpha_y M_3^{k+\mu_x} B_2^{\frac{\alpha_y\mu_y}{2}}} ((M_3 B_2)^{k+\mu_x}$$
$$\sum_{i=0}^{q-1} \frac{(-1)^i}{i!} \Gamma(-\frac{p}{q}(i+k+\mu_x)+\mu_y)(M_3 B_2)^i \frac{z^{2i+2(k+\mu_x)}}{2i+2(k+\mu_x)}$$
$$_2F_{p+q+1}(1,\frac{2i+2(k+\mu_x)}{2q};\Xi(p,\frac{p}{q}(i+k+\mu_x)-\mu_y+1),\Xi(q,i+1),\frac{2i+2(k+\mu_x)}{2q}+1;\frac{(M_3 B_2 z^2)^q}{(-p)^p(-q)^q}) + \frac{\alpha_y}{2}(M_3 B_2)^{\frac{\alpha_y\mu_y}{2}} \quad (22)$$
$$\sum_{i=0}^{p-1} \frac{(-1)^i}{i!} \Gamma(k+\mu_x-\frac{q}{p}(i+\mu_y))(M_3 B_2)^{\frac{\alpha_y i}{2}} \frac{z^{\alpha_y i+\alpha_y\mu_y}}{\alpha_y i+\alpha_y\mu_y}$$
$$_2F_{p+q+1}(1,\frac{\alpha_y i+\alpha_y\mu_y}{\alpha_y q};\Xi(q,\frac{q}{p}(i+\mu_y)-k+\mu_x+1),\Xi(p,i+1),\frac{\alpha_y i+\alpha_y\mu_y}{\alpha_y q}+1;\frac{(M_3 B_2 z^2)^{\frac{\alpha_y p}{2}}}{(-p)^p(-q)^q}))$$

---

$$\overline{C}(\gamma)|_0^{\infty} = \frac{B}{\ln 2} \sum_{n=0}^{\infty} \frac{(-1)^n}{(n+1)!} \sum_{k=0}^{\infty} \frac{A_1 M_1 M_2}{4\Gamma(k+1)\Gamma(k+\mu_x)} (\frac{\mu_x^2 \kappa_x(1+\kappa_x)}{\hat{r}_x^2})^k \frac{2}{\alpha_y M_3^{k+\mu_x} B_2^{\frac{\alpha_y\mu_y}{2}}} ((M_3 B_2)^{k+\mu_x} \sum_{i=0}^{q-1} \frac{(-1)^i}{i!} \Gamma(-\frac{p}{q}(i+k+\mu_x)+\mu_y)$$
$$(M_3 B_2)^i \frac{\gamma^{n+i+(k+\mu_x)+1}}{(n+i+(k+\mu_x)+1)\overline{\gamma}^{i+(k+\mu_x)}}$$
$$_2F_{p+q+1}(1,\frac{n+i+(k+\mu_x)+1}{q};\Xi(p,\frac{p}{q}(i+k+\mu_x)-\mu_y+1),\Xi(q,i+1),\frac{n+i+(k+\mu_x)+1}{q}+1;\frac{(M_3 B_2)^q}{(-p)^p(-q)^q}(\frac{\gamma}{\overline{\gamma}})^q) +$$
$$\frac{\alpha_y}{2}(M_3 B_2)^{\frac{\alpha_y\mu_y}{2}} \sum_{i=0}^{p-1} \frac{(-1)^i}{i!} \Gamma(k+\mu_x-\frac{q}{p}(i+\mu_y))(M_3 B_2)^{\frac{\alpha_y i}{2}} \frac{\gamma^{n+\frac{\alpha_y i+\alpha_y\mu_y}{2}+1}}{(n+\frac{\alpha_y i+\alpha_y\mu_y}{2}+1)\overline{\gamma}^{\frac{\alpha_y i+\alpha_y\mu_y}{2}}} \quad (26)$$
$$_2F_{p+q+1}(1,\frac{n+\frac{\alpha_y i+\alpha_y\mu_y}{2}+1}{\frac{\alpha_y p}{2}};\Xi(q,\frac{q}{p}(i+\mu_y)-(k+\mu_x)+1),\Xi(p,i+1),\frac{n+\frac{\alpha_y i+\alpha_y\mu_y}{2}+1}{\frac{\alpha_y p}{2}}+1;\frac{(M_3 B_2)^{\frac{\alpha_y p}{2}}}{(-p)^p(-q)^q}(\frac{\gamma}{\overline{\gamma}})^{\frac{\alpha_y p}{2}}))|_0^{\infty}$$

---

with the aid of Taylor series [27], the logarithmic function $\ln(1+\gamma)$ can be expressed as

$$\ln(1+\gamma) = \sum_{n=0}^{\infty} (-1)^n \frac{x^{n+1}}{(n+1)!} \quad (28)$$

Substituting (27) and (28) into (25), with the help of [29, Eq. (1.16.1.1)] and after several algebraic manipulations, the ECC of $Z$ can be obtained as (26).

## V. NUMERICAL RESULTS AND APPLICATION

The PDF, CDF and ECC over the composite fading distribution model

$$f_\gamma(\gamma) = \sum_{k=0}^{\infty} \frac{A_1 M_1 M_2}{4\Gamma(k+1)\Gamma(k+\mu_x)} (\frac{\mu_x^2 \kappa_x (1+\kappa_x)}{\hat{r}_x^2})^k \frac{2}{\alpha_y M_3^{k+\mu_x} B_2^{\frac{\alpha_y \mu_y}{2}}} ((M_3 B_2)^{k+\mu_x} \sum_{i=0}^{q-1} \frac{(-1)^i}{i!} \Gamma(-\frac{p}{q}(i+k+\mu_x)+\mu_y)(M_3 B_2)^i \frac{\gamma^{i+(k+\mu_x)-1}}{\bar{\gamma}^{i+(k+\mu_x)}}$$

$$_1F_{p+q}(1;\Xi(p,\frac{p}{q}(i+k+\mu_x)-\mu_y+1),\Xi(q,i+1);\frac{(M_3 B_2)^q}{(-p)^p(-q)^q}(\frac{\gamma}{\bar{\gamma}})^q) + \frac{\alpha_y}{2}(M_3 B_2)^{\frac{\alpha_y \mu_y}{2}} \sum_{i=0}^{p-1} \frac{(-1)^i}{i!} \Gamma(k+\mu_x - \frac{q}{p}(i+\mu_y))(M_3 B_2)^{\frac{\alpha_y i}{2}}$$

$$\frac{\gamma^{\frac{\alpha_y i + \alpha_y \mu_y}{2}-1}}{\bar{\gamma}^{\frac{\alpha_y i + \alpha_y \mu_y}{2}}} {_1F_{p+q}}(1;\Xi(q,\frac{q}{p}(i+\mu_y)-(k+\mu_x)+1),\Xi(p,i+1);\frac{(M_3 B_2)^{\frac{\alpha_y p}{2}}}{(-p)^p(-q)^q}(\frac{\gamma}{\bar{\gamma}})^{\frac{\alpha_y p}{2}})) \quad (27)$$

---

$\kappa$-$\mu$/$\alpha$-$\mu$ have been simulated in MATHEMATICA [32] with different values of fading parameters $\kappa_1$, $\mu_1$, $\alpha_2$ and $\mu_2$ (where we define 1 for $x$ and 2 for $y$). Unlike [8] and [12], this paper investigates the product series representation of $\kappa$-$\mu$/$\alpha$-$\mu$, so the $\alpha_y$ is the most important parameter for simulation in (15), (22), (26) and (27). Fig.1 and Fig.2 show the PDF of the composite distribution $\kappa$-$\mu$/$\alpha$-$\mu$, in Fig.1 when $\alpha_2$=2, with different value of $\kappa_1$, as $\mu_1$ and $\mu_2$ increase, more concentrated around $\bar{y}$=1 have been produced, on the other hand, if $\mu_1$ and $\mu_2$ are very low, the PDF curves are closed to the axis of ordinates. Likewise, when $\alpha_2$=2, 6, 10, we can obtain the same simulation performance for $\mu_1$ and $\mu_2$. Fig.3 presents the CDF of composite fading distribution $\kappa$-$\mu$/$\alpha$-$\mu$ when $\alpha_2$=2, 6, 10 for varied $\mu_2$, as $\alpha_2$ and $\mu_2$ increase at the same time, the CDF curves become more abruptly and converges to 1 more fast, besides $\alpha_2$ is more important than $\mu_2$ on convergence property. Furthermore, it is noteworthy that although the phenomena of singularities for series representation have been produced by the product of the fading models like [8], [9] and [12], etc, it does not affect the accuracy and convergence for PDF and CDF.

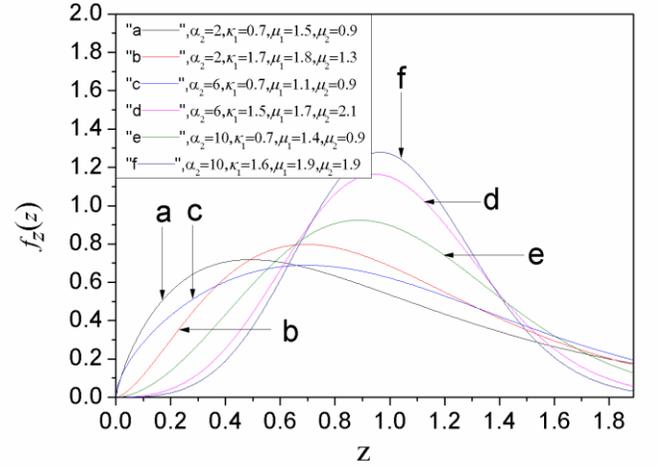

Fig.2. PDF for the composite fading distribution of $\kappa$-$\mu$/$\alpha$-$\mu$, with $\hat{r}_x = \hat{r}_y = 1$ and fading values for $\kappa_1$, $\mu_1$, $\alpha_2$ and $\mu_2$

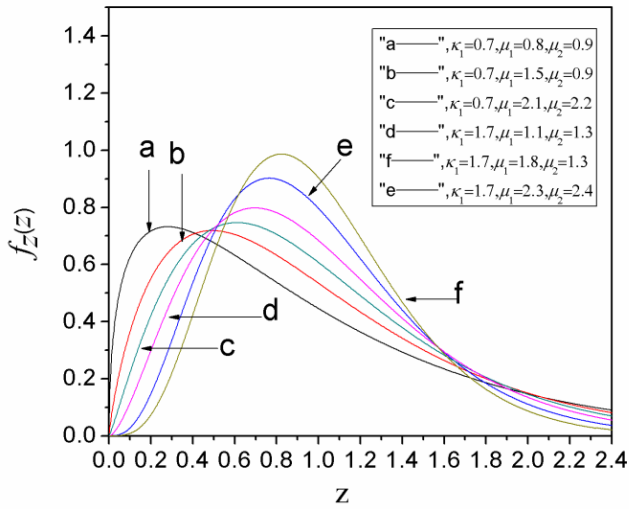

Fig.1. PDF for the composite fading distribution of $\kappa$-$\mu$/$\alpha$-$\mu$, with $\alpha_2$=2, $\hat{r}_x = \hat{r}_y = 1$ and fading values for $\kappa_1$, $\mu_1$ and $\mu_2$

Fig.4 - Fig.6 show the ECC versus average SNR for different values of fading parameters when $B=ln2$. Fig.4 and Fig.5 compare ECC versus average SNR for different values of $\kappa_1$ as $\mu_1$ and $\mu_2$ changes, when $\alpha_2$=2. It shows that larger $\kappa_1$, $\mu_1$ and $\mu_2$ can jointly improve ECC, $\mu_1$ and $\mu_2$ have more important effect than $\kappa_1$ if average SNR keeps increasing, besides when average SNR is greater than 0 dB, the increasing rate of ECC for larger $\mu_1$ and $\mu_2$ is higher. On the other hand, in Fig.6 if $\kappa_1$=0.7, $\mu_1$=1.1 and $\mu_2$=0.9, we analyze ECC versus average SNR when $\alpha_2$=2, 4, 6. It obviously indicates that under low average SNR, lower $\alpha_2$ have larger ECC, but when average SNR is greater than 2.5 dB, larger $\alpha_2$ is corresponding to higher ECC.

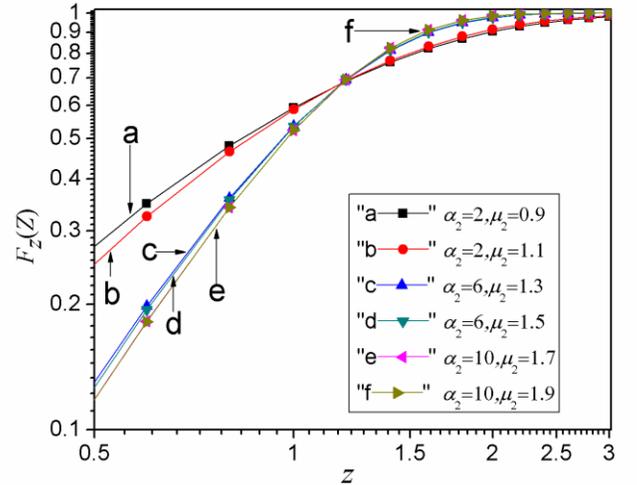

Fig.3. CDF for the composite fading distribution of $\kappa$-$\mu$/$\alpha$-$\mu$, with $\kappa_1$=1.1, $\mu_1$=1.2, $\hat{r}_x = \hat{r}_y = 1$ and fading values for $\alpha_2$ and $\mu_2$

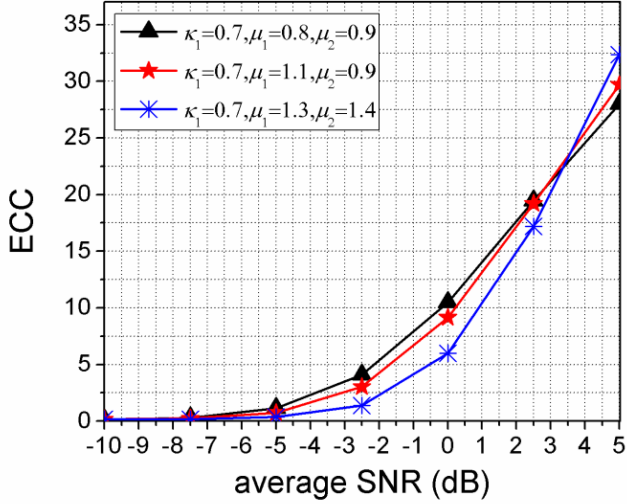

Fig.4. ECC versus average SNR for the composite fading distribution κ-μ/α-μ, with $B=\ln 2$, $α_2=2$, $\hat{r}_x = \hat{r}_y = 1$ and $κ_1=0.7$ for different values of $μ_1$ and $μ_2$

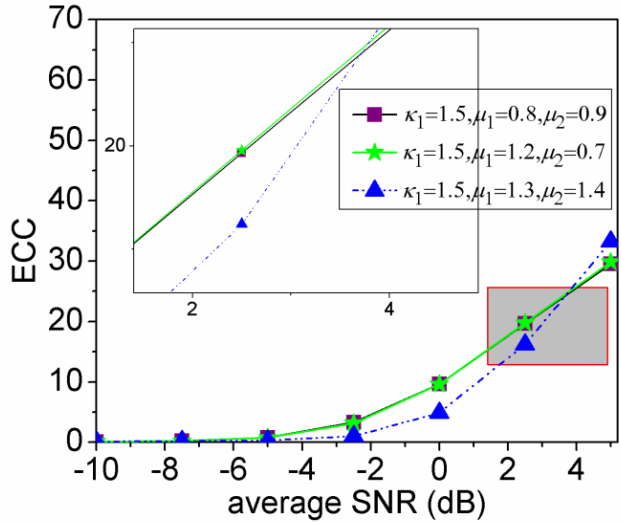

Fig.5. ECC versus average SNR for the composite fading distribution κ-μ/α-μ, with $B=\ln 2$, $α_2=2$, $\hat{r}_x = \hat{r}_y = 1$ and $κ_1=1.5$ for different values of $μ_1$ and $μ_2$

## VI. CONCLUSIONS

In this paper, the product model of two i.n.i.d. RVs for κ-μ and α-μ fading distributions have been derived, this model can improve spectrum efficiency in CRs and evaluate the performance of communication systems as vehicle-to-vehicle communications, cascaded fading channels and body area networks. The novel exact series representations for PDF and CDF of κ-μ/α-μ are derived to investigate the performance metrics of fading parameters, and novel composite fading channels such as Rice/α-μ, Rayleigh/α-μ, Nakagami-m/α-μ and one-sided Gaussian/α-μ can be obtained from the κ-μ/α-μ model as special cases, the exact series representation with generalized hypergeometric function have been derived with the method of algebra to compute more efficiently instead of Fox H-function in [12]. Furthermore, ECC over proposed generalized composite fading channels have been derived to evaluate the spectral efficiency in multiple communication scenarios with different fading parameters values. In general, the proposed κ-μ/α-μ model can be widely used as generalized composite multipath-shadowing fading scenarios in wireless communication fields.

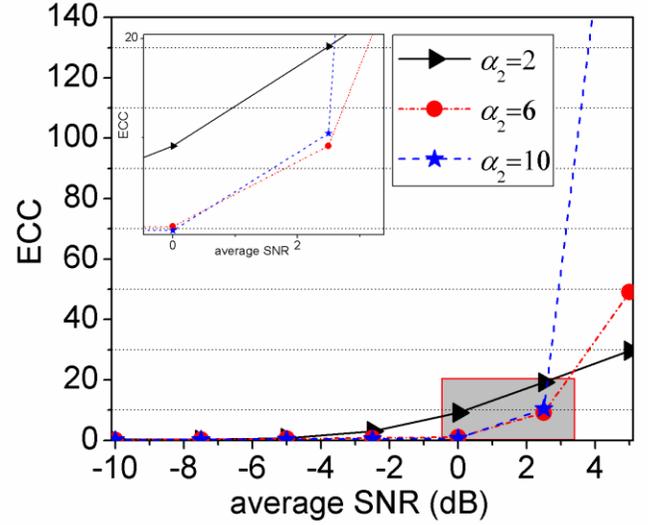

Fig.6. ECC versus average SNR for the composite fading distribution κ-μ/α-μ, with $B=\ln 2$, $\hat{r}_x = \hat{r}_y = 1$, $κ_1=0.7$, $μ_1=1.1$ and $μ_2=0.9$ for different values of $α_2$